\documentclass[preprint,conference]{IEEEtran}
\usepackage{cite}
\usepackage{amsmath,amssymb,amsfonts}
\usepackage{algorithmic}
\usepackage{graphicx}
\usepackage{textcomp}
\usepackage{placeins}
\usepackage{xcolor}
\def\BibTeX{{\rm B\kern-.05em{\sc i\kern-.025em b}\kern-.08em
    T\kern-.1667em\lower.7ex\hbox{E}\kern-.125emX}}

\usepackage{tikz}
\usepackage{textcomp}
\usepackage{hyperref}
\newcommand\copyrighttext{%
  \footnotesize \textcopyright 2022 IEEE. Personal use of this material is permitted.
  Permission from IEEE must be obtained for all other uses, in any current or future 
  media, including reprinting/republishing this material for advertising or promotional 
  purposes, creating new collective works, for resale or redistribution to servers or 
  lists, or reuse of any copyrighted component of this work in other works. 
  DOI: \href{<https://ieeexplore.ieee.org/document/9810572>}{10.1109/PMAPS53380.2022.9810572}}
\newcommand\copyrightnotice{%
\begin{tikzpicture}[remember picture,overlay]
\node[anchor=south,yshift=10pt] at (current page.south) {\fbox{\parbox{\dimexpr\textwidth-\fboxsep-\fboxrule\relax}{\copyrighttext}}};
\end{tikzpicture}%
}

\begin{document}

\title{Benchmarking Explanatory Models for Inertia Forecasting using Public Data of the Nordic Area
\\

}

\author{\IEEEauthorblockN{Jemima Graham}
\IEEEauthorblockA{ 
\textit{Imperial College London}\\
London, United Kingdom \\
jemima.graham16@imperial.ac.uk}
\and
\IEEEauthorblockN{Evelyn Heylen}
\IEEEauthorblockA{ 
\textit{Centrica Business Solutions}\\
Antwerp, Belgium  \\
evelyn.heylen@centrica.com}
\and
\IEEEauthorblockN{Yuankai Bian}
\IEEEauthorblockA{ 
\textit{National Grid ESO}\\ Wokingham, United Kingdom\\
yuankai.bian@nationalgrideso.com}
\and
\IEEEauthorblockN{Fei Teng}
\IEEEauthorblockA{ 
\textit{Imperial College London}\\
London, United Kingdom\\
f.teng@imperial.ac.uk}}


\maketitle

\copyrightnotice

\begin{abstract}
This paper investigates the performance of a day-ahead explanatory model for inertia forecasting based on field data in the Nordic system, which achieves a 43\% reduction in mean absolute percentage error (MAPE) against a state-of-the-art time-series forecast model. The generalizability of the explanatory model is verified by its consistent performance on Nordic and Great Britain datasets. Also, it appears that a long duration of training data is not required to obtain accurate results with this model, but taking a more spatially granular approach reduces the MAPE by 3.6\%. Finally, two further model enhancements are studied considering the specific features in Nordic system: (i) a monthly interaction variable applied to the day-ahead national demand forecast feature, reducing the MAPE by up to 18\%; and (ii) a feature based on the inertia from hydropower, although this has a negligible impact. The field dataset used for benchmarking is also made publicly available.
\end{abstract}


\maketitle


\section{Introduction}
\label{sec:introduction}

Inertia levels in power systems have been decreasing over the last decade causing operational challenges. Traditionally, inertia was abundant and originated from the rotation of masses in synchronous generators and motor loads; however, in recent years there has been an influx of renewable energy sources (RES) that are connected to the grid asynchronously (i.e. via power electronics). As these power sources do not deliver inertial response naturally, power systems are transitioning to low inertia systems \cite{Heylen2021, NGESO2019, Milano2018}. Low inertia systems are at particular risk of frequency instabilities due to power imbalances. The reason for this is that the rate of change of frequency is higher in such systems which leaves the system less time to react to the imbalances \cite{Milano2018, Du2018}. As a result, having sufficient inertia in power systems at each point in time is essential to ensure that power system operation is secure and reliable \cite{Heylen2021}. For this reason, system operators need to accurately forecast the amount of inertia expected in the system to aid their decision-making for frequency response management \cite{Heylen2021, Matevosyan2018}.

While this was not necessary in the past due to the consistent usage of synchronous generators \cite{Teng2016}; in low-inertia systems with high RES penetration, the number of synchronous generators required will vary greatly over time as RES are often weather-dependent \cite{Teng2016}. This variation will increase as more RES are introduced into energy systems in accordance with the push towards carbon-neutrality. For example, in the UK, the government aims to have 40 GW of offshore wind by 2030 \cite{UKgovt2020}. This offshore wind will further displace traditional synchronous generators and will introduce more uncertainty into inertia forecasts. Contrastingly, in the Nordic, this is less likely to be the case as their main RES is hydropower which still produces some inertia. 

Nevertheless, even if some energy systems are less likely to be exposed to inertia variability than others, it is still necessary for system operators (SOs) to understand the uncertainty surrounding a forecast. SOs are risk-averse due to the large-scale disruption system mismanagement could cause. For this reason, a conservative approach must be taken when forecasting inertia.

Despite the need for accurate inertia forecasts, existing work in the field of inertia forecasting has been limited so far \cite{Heylen2021}. Although forecast models have been presented in the literature \cite{Du2018, GEDigital2018, Wilson2018}, the authors are not aware of existing benchmark studies that objectively compare inertia forecast models on a publicly available dataset. Benchmarking, where models are compared to a validated base case, is crucial to meaningfully assess the performance of any newly developed inertia forecast models.

To facilitate objective comparisons of newly-developed day-ahead inertia forecast models with the state-of-the-art in the field, we developed a benchmarking methodology and benchmarking dataset in this paper. A case study is presented comparing two existing forecast models for which public datasets are available: an explanatory model we previously developed on a dataset of Great Britain (GB) \cite{Heylen2021a}; and a time-series forecast model developed on a dataset of the Nordic area \cite{GonzalezLongatt2020}. The performance of the two models is compared on the Nordic field dataset. Additionally, this study evaluates the generalizability of the explanatory model by applying it to datasets of both the Nordic and GB power systems. 



These benchmarking efforts are coupled with a two-pronged investigation into the characteristics of the model: the first branch of the investigation explores the spatial and temporal dependencies of the model; and the second branch of the investigation considers whether the model can be further developed by: (i) considering the annual seasonality of the inertia through a monthly interaction variable; and (ii) considering whether a large amount of hydropower in the generation mix requires specific adaptations in the explanatory model. In addition to these investigations, this study examines the impact of explanatory variable forecast errors on the accuracy of the inertia forecast model.


The remainder of this paper is organized as follows: Section \ref{sec:methodology} describes the benchmarking methodology; Section \ref{sec:case_study} describes the Nordic dataset used to validate and test the model; Section \ref{sec:results&discussion} discusses the results of this study; and Section \ref{sec:conclusion} considers any concluding remarks. 

\section{Methodology}
\label{sec:methodology}

This section introduces the explanatory and time-series forecast models that are compared in this study. 

\subsection{Explanatory and Time-series Inertia Forecast Models}
\label{subsec:benchmarking}
\FloatBarrier

The explanatory inertia forecast model is given in \eqref{eq:evelyn}:
\begin{align}
\label{eq:evelyn}
     \hat{E}^{I,G}_{t+k|t} = & \alpha_{d,1}E^{I,G}_{t} + \alpha_{d,2}\hat{P}^{ND}_t + \alpha_{d,3}\hat{P}^{\textrm{wind}}_t \\ & + \alpha_{d,4}\hat{P}^{\textrm{solar}}_t + \alpha_{d,5}P^{IC}_t + \alpha_{d,6}t + \alpha_{d,7}t^2 \nonumber 
\end{align}
where $E^{I,G}_t$ is the kinetic energy (inertial energy) of the system at time $t$; $\hat{P}^{ND}_t$ is the day-ahead national demand forecast at time $t$; $\hat{P}^{\textrm{wind}}_t$ is the day-ahead wind power forecast at time $t$; $\hat{P}^{\textrm{solar}}_t$ is the day-ahead solar power forecast at time $t$; $P^{IC}_t$ is the interconnection flow at time $t$; and $\alpha_{d,i}$ are coefficients where $d$ is whether it is a weekday or weekend/holiday \cite{Heylen2021a}. A detailed discussion of the development of this model can be found in \cite{Heylen2021a}. This work was validated on a publicly available GB dataset. As the data was half-hourly, data from 2016 and 2017 made up the training set and data from 2018 was used for the test set. Overall, the model obtains good results on this dataset, with a mean absolute percentage error (MAPE) of 4.2\% \cite{Heylen2021a}.

In the past, the error bound on inertia measurements and forecasts has been unclear. To capture the uncertainty of the inertia forecast model described by (\ref{eq:evelyn}), the following Gaussian distribution can be assumed \cite{Heylen2021a}:
\begin{equation}
    \hat{F}_{t+k|t}(E^{I,G}_{t+k};\textbf{x}_t) = \Phi(E^{I,G}_{t+k}; \hat{\mu}_{t+k|t}, \hat{\sigma})
\end{equation}
where $\Phi(\cdot)$ denotes the Gaussian distribution; the mean $\hat{\mu}_{t+k|t}$ is equal to $\hat{E}^{I,G}_{t+k|t}$; and standard deviation $\hat{\sigma}$ is assumed to be a constant equal to the sample standard deviation of the training data \cite{Heylen2021a}:
\begin{equation}
    \hat{\sigma} = \sqrt{\frac{\sum_{t}(E^{I,G}_{t} - \mu)^2}{N}}
\end{equation}
where $\mu$ is the mean of the training data; and $N$ is the number of samples in the training set.

Contrastingly, the time-series forecast model is given in \eqref{eq:gonzlez}:
\begin{align}
\label{eq:gonzlez}
    \hat{E}^{I,G}_{t} = g_t + s_t + h_t
\end{align}
where $g_t$ is the trend component, defined as a logistic growth model; $s_t$ is the seasonality component, defined as a Fourier series; and $h_t$ is an irregular component that describes any holidays or special events. A detailed discussion of the development of this model can be found in \cite{GonzalezLongatt2020}. This work was validated on a publicly available Nordic dataset. This dataset included data with minutely resolution. As a result, the training and test sets covered a shorter duration of time; the training set in this study spanned 1st - 30th January 2018 and the test set was 31st January 2018. This model is used to generate short-term forecasts which predict inertia at least one hour into the future and at most twenty-four hours into the future. Twenty-four hours into the future, this model achieves a MAPE of 7\% \cite{GonzalezLongatt2020}, which will be directly compared to the MAPE obtained when the explanatory model is applied to the Nordic. In order to provide a direct comparison with the work of Gonzlez-Longatt et al., the 31st January 2018 is used as a test set in this portion of the study; however, a larger training set containing one year of data spanning 31st January 2017 to 30th January 2018 is used instead.

In addition, the generalizability of the explanatory model is investigated in order to understand the applicability of the model to other power systems. This section of the work will compare the accuracy of forecasts generated using the explanatory model on both the GB and Nordic case studies. 

\subsection{Adaptation of Explanatory Model for Nordic System}
\label{subsec:benchmarking}
\FloatBarrier
Alongside the aforementioned investigations, this paper studies the potential adaption of the explanatory model to the Nordic system. In particular, the spatial and temporal dependencies of the explanatory model are explored. The temporal dependency of the model is evaluated by training the model with different durations of training data. In this part of the study, 1st January 2020 - 31st August 2020 is used as the test set, while data from 2016 - 2019 is used for the various training sets. 

The impact of spatial granularity is also examined as it is particularly relevant in the case of the Nordic where the area is comprised of four different countries. As the aim of this work is to forecast inertia for the whole of the Nordic, this approach involves forecasting inertia by region and aggregating these results to produce a forecast for the whole of the Nordic. This model has the following form:
\begin{equation}
    \hat{E}^{I,G}_{t+k|t} = {\sum_r} \hat{E}^{I,G}_{r, t+k|t}
    \label{eq:spatial_granularity1}
\end{equation}
where each of the $\hat{E}^{I,G}_{t+k|t}$ are forecasted according to \eqref{eq:evelyn} and $r$ indicates the region. 

Furthermore, potential for further development of the explanatory model is considered in two ways: (i) the incorporation of a monthly interaction variable; and (ii) the inclusion of a feature based on the inertia from hydropower. The monthly interaction variable aims to capture the annual seasonality of the inertia. Some inspiration is taken from the work of Mirasgedis et al. \cite{Mirasgedis2007} which highlights the importance of considering monthly periodicities when modelling electricity demand. As a result, both applying the monthly interaction variables to all explanatory variables, and applying the monthly interaction variable only to the day-ahead national demand forecast variable is trialled. Additionally, the feature based on inertia from hydropower the previous day allows the impact of having large amounts of hydropower in the generation mix to be considered. This feature is incorporated into the existing explanatory model and the impact on forecast accuracy is evaluated.


Finally, this paper investigates how much the inertia forecast accuracy can be improved by improving the underlying day-ahead forecasts. In order to conduct this investigation, the real-time values for the national demand, wind power, and solar power are collected to mimic a forecast with 100\% accuracy. These real-time values are each substituted into the explanatory model so that their individual impacts can be ascertained. The overall impact from having all three forecasts as 100\% accurate forecasts is also examined. 



\section{The Nordic dataset}
\label{sec:case_study}

The Nordic power system is chosen as a test case due to the abundance of publicly available data from the Nordic TSOs. Publicly available datasets that are verified by TSOs are difficult to find, especially one that contains all of the explanatory variables required for \eqref{eq:evelyn} in Section \ref{sec:methodology}. While the publicly available GB dataset used by Heylen et al. when developing \eqref{eq:evelyn} does contain all of the relevant explanatory variables, it has not been verified by National Grid ESO. 

Regardless, the power system in the Nordic shares some similarities with the GB power system. 
In particular, both systems are of a similar size, with a peak load of 60-70 GW and minimum load of 20-25 GW \cite{Tilastokeskus2016}. 
However, a key difference between the Nordic and GB power systems is the energy generation mix. In the Nordic, the energy generation mix is predominantly hydropower, followed by nuclear and wind. Contrastingly, in GB, over the beginning of the dataset used here, the energy generation mix is predominantly natural gas followed by nuclear and wind. This transitions to a more wind dominant energy system as we look towards present day with wind accounting for up to 50\% of the energy generation mix. As a large proportion of hydropower is seen in the Nordic power system and not in the GB power system, the use of inertia from hydropower as an explanatory variable is explored in this work. 

\begin{figure}
    \centering
    \includegraphics[width=0.5\textwidth]{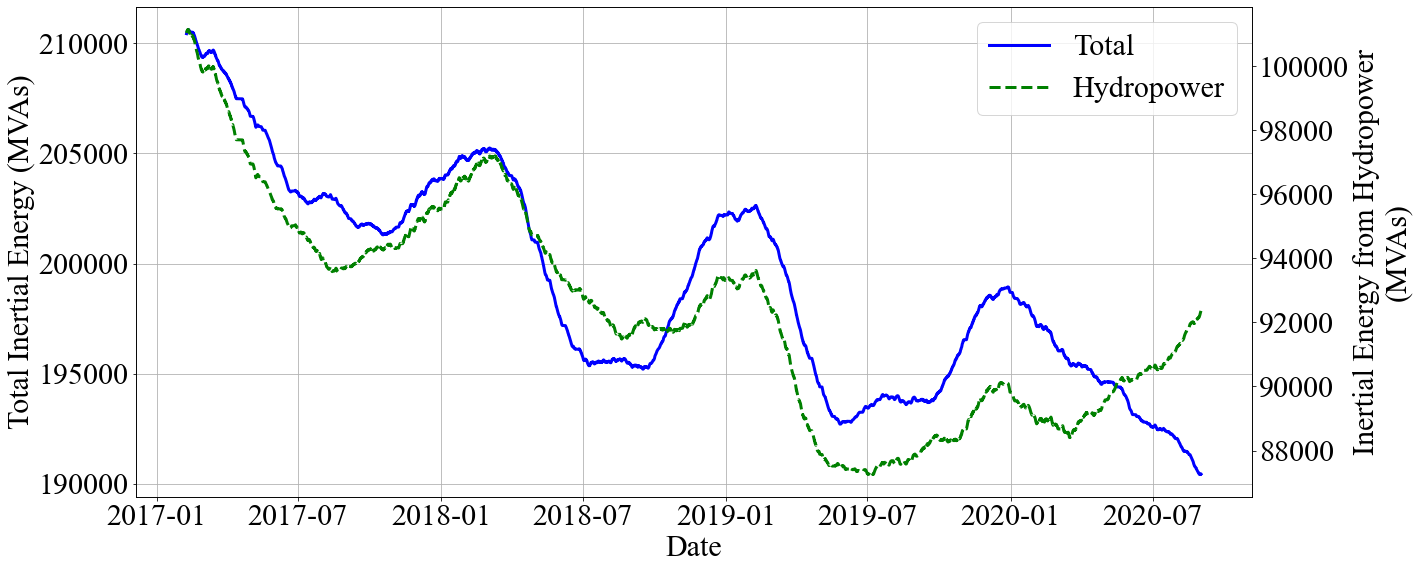}
    \caption{Trend of the total inertia (blue) and the inertia from hydropower (green) in the Nordic.}
    \label{fig:trend}
\end{figure}

\begin{figure}
    \centering
    \includegraphics[width=0.5\textwidth]{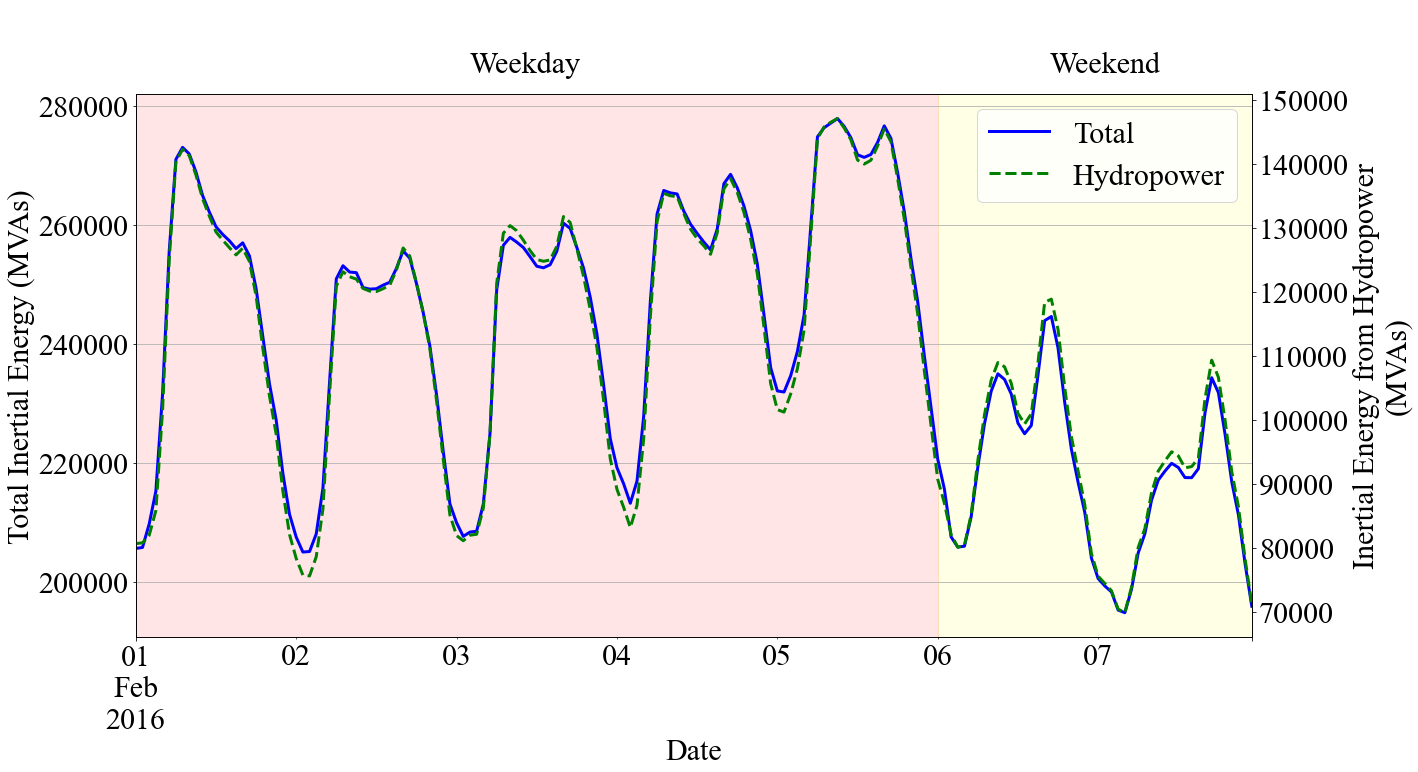}
    \caption{Total inertia and inertia from hydropower in the Nordic during the week beginning 1st February 2016. The portion of the graph falling on a weekday is shaded in red and the the portion of the graph falling on a weekend is shaded in yellow.}
    \label{fig:periodicity}
\end{figure}

In addition to the general characteristics of the Nordic power system, it is important to understand the inertial energy characteristics of the region so that this can be contrasted with other regions in future work. 
From Fig.\ref{fig:trend}, it can be seen that the amount of inertial energy in the Nordic is decreasing over time. This behaviour is likely due to the increased share of RES in the power system \cite{Wilson2019}.  Multiple seasonalities also exist; not only is there an annual seasonality (as shown in Fig. \ref{fig:trend}), there is also a daily seasonality which can be seen in Fig.\ref{fig:periodicity}. These periodicities led the authors to consider a monthly interaction variable which will be discussed in greater depth in Section \ref{sec:results&discussion}. Additionally, it is clear from Fig.\ref{fig:periodicity} that whether it is a weekend or weekday affects the quantity of inertia in the system. This was part of the motivation behind including an interaction variable based on whether it was a weekday or a weekend/holiday in (\ref{eq:evelyn}) \cite{Heylen2021a}.

Given that hydropower makes up a large proportion of the energy generation mix in the Nordic, it is important to consider the characteristics of the inertia from hydropower. Fig.\ref{fig:trend} illustrates that the inertial energy from hydropower decreases over time in a similar way to the total inertial energy. However, it does appear to increase towards the end of the dataset, perhaps indicating a stabilization in the quantity of inertia from hydropower. Additionally, the inertial energy from hydropower displays similar periodicity patterns to the total inertial energy as shown in Fig.\ref{fig:trend} and Fig.\ref{fig:periodicity}.




 In the dataset collected under this study, three features are day-ahead forecasts meaning that there is some predictive error associated with these values. As discussed in Section \ref{sec:methodology}, the impact of this will be investigated by replacing the forecasts with real-time values. The authors considered that this may have an impact: a MAPE of approximately 1\% is found for the national demand forecast; a symmetric MAPE (sMAPE) of approximately 38\% is found for the solar power forecast; and a MAPE of 14\% is found for the wind power forecast.

Overall, the data used in this case study spans January 2016 to August 2020. It has hourly resolution in contrast to the minutely resolution used in time-series forecasting studies in \cite{GonzalezLongatt2020}. Consequently, in order to provide a direct comparison with the work in \cite{GonzalezLongatt2020}, a test set of 31st January 2018 will be used with a training set between 31st January 2017 and 30th January 2018. For all other investigations, the test set will be between 1st January 2020 and 31st August 2020, and all other data will be used as the training set.



\section{Results \& Discussion}
\label{sec:results&discussion}

This section covers three topics: benchmarking of the explanatory model developed in \cite{Heylen2021a} (Section \ref{subsec:benchmarking}); spatial-temporal dependencies of the model (Section \ref{subsec:spatial-temporal}); and additional variables developed to further improve model accuracy (Section \ref{subsec:additional_variables}). It must be noted that we consider a substantial improvement in model accuracy to be a difference of 1000 MVAs or more as this will cause a notable change in ESO planning.

\subsection{Benchmarking}
\label{subsec:benchmarking}

Benchmarking of the explanatory model was carried out in two ways: (i) the model was compared against the state-of-the-art time-series forecast model developed in \cite{GonzalezLongatt2020}; and (ii) the model performance on the Nordic dataset is compared to the model performance on the GB dataset. These results are given in Table \ref{tab:gonzalez-longatt} and Table \ref{tab:casestudy_comparison} respectively. 

Table \ref{tab:gonzalez-longatt} emphasizes that the explanatory model developed in \cite{Heylen2021a} outperforms the time-series forecast model developed in \cite{GonzalezLongatt2020}, reducing the MAPE by 43\% (approximately 86000 MVAs). This implies that considering a variety of relevant variables such as: day-ahead national demand forecast; day-ahead wind power forecast; day-ahead solar power forecast; and interconnection flow; improves the accuracy of inertia forecasts. These results also set a benchmark of 4\% MAPE for future inertia forecast development on the Nordic case study. 

\begin{table}[ht!]
    \centering
    \caption{Test MAPEs for different day-ahead inertia forecasts.}
    \begin{tabular}{|p{3.5cm}|p{3cm}|}
    \hline
      \textbf{Model} &  \textbf{Day-Ahead Forecast Test MAPE (\%)}\\
      \hline
       Gonzalez-Longatt et al. \cite{GonzalezLongatt2020}  & 7 \\
       \hline
       Heylen et al. \cite{Heylen2021a} & 4 \\
       \hline
    \end{tabular}
    \label{tab:gonzalez-longatt}
\end{table}

In addition, Table \ref{tab:casestudy_comparison} indicates that the accuracy of the explanatory model developed in \cite{Heylen2021a} is consistent; there is little variation in training or test MAPE between the GB and Nordic case studies despite the differences between the case studies outlined in Section \ref{sec:case_study}. This suggests that even though this explanatory model was developed for use on a GB dataset, it is generalizable to other power systems. One important point to note is that there are some key similarities between the GB and Nordic power systems in terms of size and composition. For this reason, application of this explanatory model to a system with a different size and composition may be useful to quantify the generalizability of this model.

\begin{table}[ht!]
    \centering
    \caption{Training and test MAPEs for the explanatory model trained on two years of data for different case studies}
    \begin{tabular}{|p{2cm}|p{2cm}|p{2cm}|}
        \hline
        \textbf{Case study} & \textbf{Training MAPE (\%)} & \textbf{Test MAPE (\%)} \\
        \hline
         Great Britain & 4.2 & 4.7 \\
         \hline
         Nordic & 4.3 & 4.5 \\
         \hline
    \end{tabular}
    \label{tab:casestudy_comparison}
\end{table}


\subsection{Spatial-temporal dependencies}
\label{subsec:spatial-temporal}

\begin{figure*}
    \centering
    \includegraphics[width=0.9\textwidth]{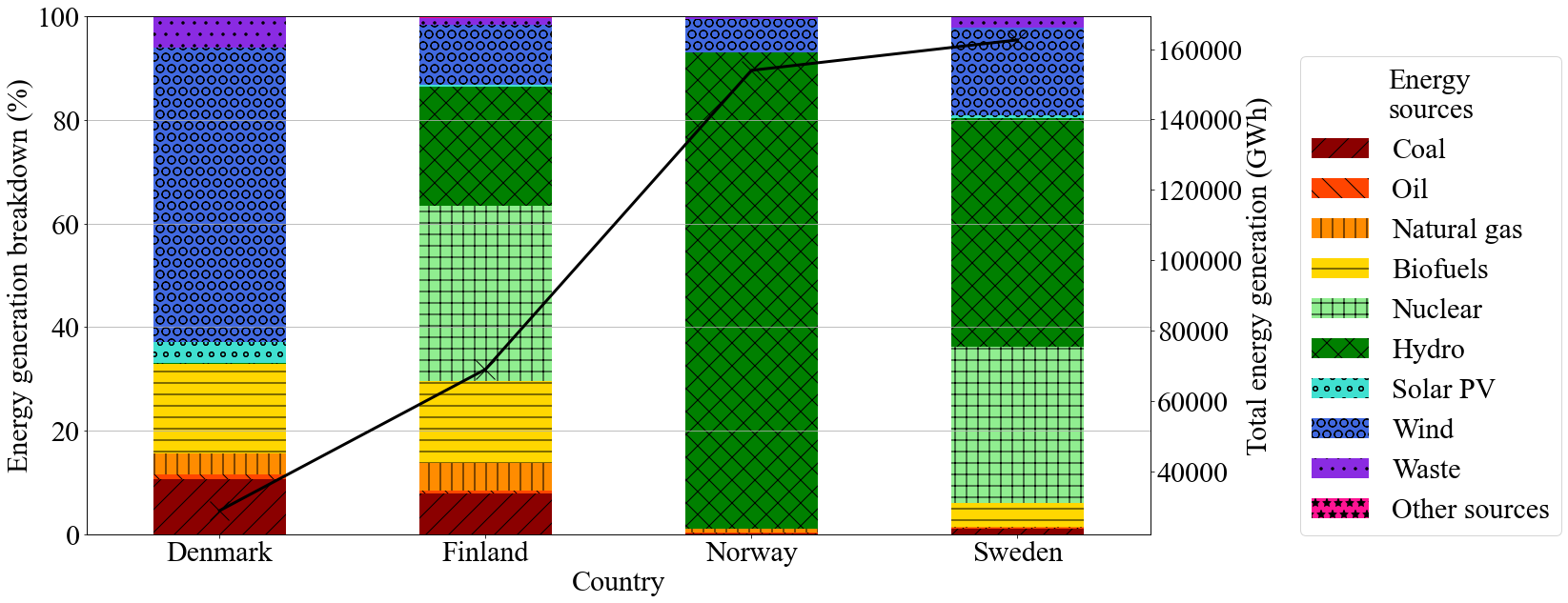}
    \caption{Energy generation of each Nordic country in 2020 as a percentage of total generation. Total energy generation in GWh is shown by the black line. Data obtained from \cite{IEA2022}.}
    \label{fig:energy_gen}
\end{figure*}

Spatial-temporal impacts are considered in two ways: (i) the impact of using a longer duration of training data is investigated by training the model using different amounts of training data; and (ii) the impact of spatial granularity on the explanatory model is investigated by treating each of the regions within the Nordic (Eastern Denmark, Finland, Norway, and Sweden) separately as outlined in Section \ref{sec:methodology}. 

Table \ref{tab:training_set_lengths} shows the training and test MAPEs for the explanatory model trained with different durations of training data. As these results show no clear trend, it implies that the duration of training data has little impact on the accuracy of the inertia forecast. The best performing explanatory model by test MAPE was the model trained using only one year of data. This model will be used as the base case going forward.

\begin{table}[ht!]
    \centering
    \caption{Training and test MAPEs for forecasts with different training set lengths.}
    \begin{tabular}{|p{2.5cm}|p{2cm}|p{2cm}|}
    \hline
      \textbf{Training Set Length} & \textbf{Training MAPE (\%)} & \textbf{Test MAPE (\%)} \\
    \hline
      1 year & 4.539 & 4.420 \\
      \hline
      2 years & 4.320 & 4.505 \\
      \hline
      3 years & 4.197 & 4.862 \\
      \hline
      4 years & 4.426 & 4.553 \\
      \hline
    \end{tabular}
    \label{tab:training_set_lengths}
\end{table} 

The more spatially granular model outperforms the base case; while the base case has a test MAPE of 4.420\%, this model achieves a MAPE of 4.261\%, which is a 3.6\% reduction (equivalent to approximately 7000 MVAs). Additionally, the training MAPE reduces by 2.6\% from 4.539\% to 4.420\% (equivalent to approximately 5000 MVAs). The main reason for this improvement is that the energy generation characteristics differ between Nordic regions both in terms of composition and quantity, as shown in Fig.\ref{fig:energy_gen}. For example, Denmark proportionally relies much more on wind and solar power compared to the other Nordic countries. 

Conversely, this improvement was found to be unrelated to tailoring the weekday/weekend and holiday interaction variable to the region. In the model that considers the Nordic as a whole, only national holidays that are common among all of the regions are considered; however, during the spatial granularity investigation, holidays specific to each of the four regions were considered. This change in holiday definition had a negligible impact on model performance, likely due to the fact that the major (and therefore, most impactful) holidays are common among all regions. Altogether, spatial granularity has a notably positive impact on the model due to the increased ability to account for regional energy generation characteristics and is recommended for use in future models.


\subsection{Additional variables}
\label{subsec:additional_variables}

\begin{figure*}[]
    \centering
    \includegraphics[width=\textwidth]{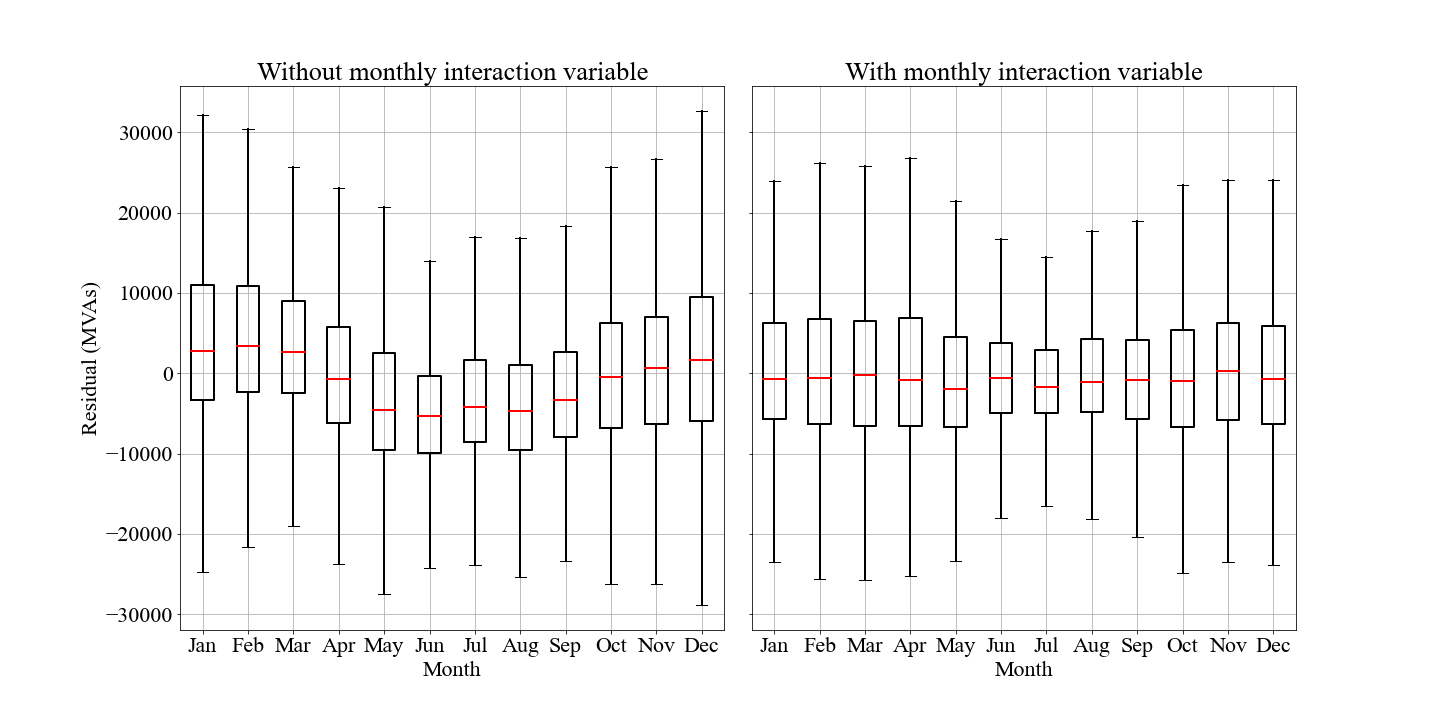}
    \caption{Comparison of residuals obtained by models both with and without the monthly interaction variable on the day-ahead national demand forecast. Both models are trained on two years of training data, and the residuals are shown for the training dataset to show the variation over a full calendar year.}
    \label{fig:month_boxplots}
\end{figure*}

Two model developments were trialled as part of this section: (i) the introduction of a monthly interaction variable on the day-ahead national demand forecast feature; and (ii) the introduction of a feature that considers inertia from hydropower the previous day. Additionally, the impact of errors from the day-ahead forecast feature errors are considered.

As discussed in Section \ref{sec:methodology}, a monthly interaction variable is applied to the day-ahead national demand forecast feature in order to better model the annual seasonality in this feature. The authors expected that this feature would be heavily influenced by month due to the work of Mirasgedis et al. \cite{Mirasgedis2007}. The results of this investigation can be seen in Table \ref{tab:monthly_interaction} and Fig.\ref{fig:month_boxplots}. In Table \ref{tab:monthly_interaction}, the MAPEs of all forecasts improve apart from the forecast that is trained on only one year of training data. This suggests that the monthly interaction variable needs at least 2 years worth of data to be well tuned. Additionally, it can be seen in Fig.\ref{fig:month_boxplots} that the variation in the mean and spread of the residuals reduces between different months if a monthly interaction feature is applied to the day-ahead national demand forecast.

Conversely, this improvement in MAPE was not seen if the monthly interaction variable is applied to all features or just the feature based on inertial energy from the previous day. This suggests that the day-ahead national demand forecast has a monthly relationship that is not seen in the other features. Overall, applying a monthly interaction variable to the day-ahead national demand forecast feature improves the model accuracy provided that at least two years of training data is used.

\begin{table}[ht!]
    \centering
    \caption{Monthly interaction variable applied to day-ahead national demand forecast feature for varying training set lengths.}
    \begin{tabular}{|p{2cm}|p{2cm}|p{2cm}|}
    \hline
       \textbf{Training Set Length} & \textbf{Training MAPE (\%)} & \textbf{Test MAPE (\%)}  \\
       \hline
      1 year & 3.553 & 7.700 \\
      \hline
      2 years & 3.600 & 3.870 \\
      \hline
      3 years & 3.447 & 4.572 \\
      \hline
      4 years & 3.679 & 4.115 \\
      \hline
    \end{tabular}
    \label{tab:monthly_interaction}
\end{table}


In addition, the authors believed that using inertia from hydropower the previous day as a feature may improve the inertia forecast model accuracy because the majority of generation in the Nordic comes from hydropower. Despite this, using the inertial energy from hydropower the previous day as a feature seems to have a negligible impact. A potential reason for this behaviour could be that the contribution of the inertia from hydropower feature is already considered by using the feature of inertia from the previous day.

Finally, the impact of the day-ahead forecast feature errors was considered in order to ascertain whether model accuracy could be improved by improving the accuracy of the day-ahead national demand forecast, the day-ahead wind power forecast, or the day-ahead solar power forecast, which are used as features in the explanatory model. It was found that the difference between the base case and the scenario with all forecasts replaced with real-time values is only around 100 MVAs which is not large enough to affect operational decisions significantly. Therefore, while the feature forecast errors slightly reduce inertia forecast accuracy, it is considered to be a negligible effect.

\section{Conclusion}
\label{sec:conclusion}

Altogether, the day-ahead explanatory inertia forecast model is applied under this study, which demonstrates good performance compared to state-of-the-art inertia forecasting techniques based on time-series forecasting. It also shows similar performance on both the GB and Nordic datasets, implying that this model is transferable to other regions. 
Consequently, the explanatory inertia forecast model discussed here is a suitable benchmark for future inertia forecast development projects on the Nordic and GB case studies.

In addition, the explanatory model was found to have limited dependence on the duration covered by the training dataset, but significant dependence on spatial granularity. The former highlights that when using this model, there is no need for especially long durations of training data, and the latter suggests that a more spatially granular approach is beneficial to model accuracy. The fact that a long duration of training data is not essential for good model performance will make it easier to collect further case studies on which this model can be trialled. 

Another key finding of this work was that introducing a monthly interaction variable on the day-ahead national demand forecast feature notably improves the accuracy of the inertia forecast. Consequently, this, alongside the spatially granular modelling approach, is recommended for use in future inertia forecast models, particularly in the probabilistic inertia forecast model developed in \cite{Heylen2021a}.

\section*{Acknowledgements}

This work has been funded by National Grid ESO under Electricity Network Innovation
Allowance project ``Short-term System Inertia Forecast"
(NIA-NGSO0020) and by EPSRC under Grant EP/R513052/1. The authors would like to thank Mr Mikko Kuivaniemi from Fingrid, Finland who contributed the data on inertial energy by country and inertial energy from hydropower both for the whole region and the Nordic countries individually. The dataset used for benchmarking can be accessed on zenodo.org (DOI: 10.5281/zenodo.5655048), and used under the Creative Commons Attribution Licence (CC BY).

\bibliographystyle{unsrt}
\bibliography{Nordic_conference_paper.bib}

\end{document}